  \documentclass[prl,aps,twocolumn,showpacs]{revtex4}
\usepackage[dvips]{graphicx}
\usepackage{dcolumn}
\newcommand{\be}{\begin{equation}}
\newcommand{\ee}{\end{equation}}
\newcommand{\bea}{\begin{eqnarray}}
\newcommand{\eea}{\end{eqnarray}}
\newcommand{\nn}{ \nonumber}
\newcommand{\ds}{\displaystyle}
\begin{document}
\topmargin=-5mm

\title {\bf Quantum Oscillations of Elastic Modiuli and 
Softening of Phonon Modes in Metals}

 \vspace{0mm}

\author{ Natalya A. Zimbovskaya$^{a,b}$ }
\affiliation{
 $^a$Department of Physics \& Astronomy,
St. Cloud State University,
720 Fourth Avenue South, St. Cloud, MN 56301,  
$^b$Department of Physics and Electronics, 
University of Puerto Rico at Humacao, 
CUH Station, Humacao, PR, 00791} 

 \begin{abstract} 
 In this paper we present a theoretical analysis of the 
effect of magnetostriction on quantum oscillations of 
elastic constants
in metals under strong magnetic fields. It is shown that at 
low temperatures a significant softening of some acoustic 
modes could occur near peaks of quantum oscillations of the 
electron density of states (DOS) at the Fermi surface (FS). 
This effect is caused by the Condon magnetic instability, 
and it can give rise to a lattice instability. We 
show that the most favorable conditions 
for this instability to be revealed occur in metals whose 
Fermi surfaces include nearly cylindrical segments.
  \end{abstract}

\date{\today} 
\maketitle

Experimental data concerning quantum oscillations in various 
observables in metals under strong magnetic fields were 
repeatedly used in studies of their electron characteristics. 
At low temperatures when the parameter $\theta $ is small 
compared to unity $ (\theta = 2\pi^2 T /\hbar \Omega; \
T $ is the temperature expressed in units of energy; $ \hbar 
\Omega $ is the value of a cyclotron quantum) these 
oscillations can  exhibit a rich structure. The latter could 
be analyzed even within a simple isotropic model.
For instance, it is known that the longitudinal magnetic 
susceptibility of a metal can reveal divergencies at peaks of
 quantum oscillations at low temperatures [1--3]. These 
divergencies crucially depend on interaction among conduction 
electrons, and they indicate a possibility of a diamagnetic
phase transition in a metal, producing Condon domain structure 
near the oscillation peaks [4]. 
Also, it was proposed in some earlier works that the above 
features in the magnetic susceptibility could give rise to a 
sensible reduction of elastic moduli in metals [1,4--6]. Here, we 
demonstrate that the effect of softening of phonon modes 
near the peaks of magnetic quantum oscillations of the 
electron DOS could be significantly strengthened, and even 
the relevant structural phase transitions could occur and 
be observed under feasible experimental conditions in those 
metals whose Fermi surfaces insert nearly cylindrical belts.

At first, we derive expressions for electron contributions to 
the elastic constants. To simplify further calculations we 
assume that the FS is axially symmetric, and the external 
magnetic field $ \bf B $ is directed along the symmetry axis 
("z" axis of the chosen coordinate system). We   analyze the 
elastic response of a metal to an external deformation 
described with the lattice displacement vector $ \bf u (r).$ 
The effect of conduction electrons on the crystalline lattice 
arises due to a self-consistent electric field which appears 
under deformation. Also, the lattice deformation gives rise 
to an additional inhomogeneous magnetic field $ \bf b(r),$ 
and to deformation induced corrections to the crystalline 
fields. Here, we omit the latter at first steps of our
analysis 
to include them later. The emergence of the electric and 
magnetic fields accompanying the lattice deformation,       
leads to a redistribution of the electron density $ N .$ The
local change in the electronic density $ \delta N \bf (r) $
 equals:
      \bea
\delta N ({\bf r})& = & - \frac{\partial N}{\partial \zeta}
e \varphi ({\bf  r}) +
\frac{\partial N}{\partial {B}}
{\bf b}({\bf  r})
 \nn \\
 & \equiv & - N_\zeta^*
\left (e \varphi ({\bf r}) +
\frac{\partial \zeta }{\partial {B}}
{\bf  b}({\bf r}) \right ) 
                               \eea
 where $ e $ is the absolute value of the electron charge. 

 The magnetic field {\bf b}({\bf r}) satisfies the equation:
      \bea  && 
\mbox {curl} {\bf  b}({\bf  r}) =
4 \pi \mbox {curl} \delta {\bf  M}({\bf  r})
    \nn \\
& = & 4 \pi \mbox {curl} \left (
\frac{\partial {\bf  M}}{\partial \zeta}
e \varphi ({\bf  r}) +
\frac{\partial {\bf M}}{\partial { \bf B}}
{\bf  b}({ \bf  r}) \right )
{\mbox {div}} {\bf  b}({\bf r}) = 0 .
      \eea
 Here { \bf M } is the magnetization vector; $ \zeta $ is the 
chemical potential of charge carriers; $ \varphi 
({\bf  r}) $ is the potential of the electrical field, 
arising due to the deformation. The quantity $ N_\zeta^*$ included into Eq.(1) is closely related to the electron density of states (DOS) on the Fermi surface $N_\zeta.$ The diffference between the two originates from the correlations in the electron system. Within the framework of the phenomenological Fermi-liquid theory the renormalized DOS $ N_\zeta^*$ has the form (See [1]): 
         \be
N_\zeta^* = - \sum \limits_{\nu \nu'} 
\frac{f_\nu - f_{\nu'}}{E_\nu -
E_{\nu'}} n_{\nu\nu'}^*{\bf (-q)} n_{\nu'\nu}{\bf (q)} 
\bigg |_{q \to 0} 
            \ee
   where $ f_\nu$ is the Fermi distribution function for 
quasiparticles with energies $ E_\nu ,$ and $ n_{\nu'\nu} 
\bf (q)$ is the Fourier transform of the operator of electron 
density in space variables. The renormalized operator of 
electron density $ n_{\nu \nu'}^* \bf(-q) $ is related to 
the ''bare'' operator $ n_{\nu\nu'} \bf (-q)$ as follows:
          \be 
n_{\nu\nu'}^* {\bf (-q)} = n_{\nu \nu'} {\bf (-q)} 
+ \sum \limits_{\nu_1
\nu_2} \frac{f_{\nu_1} - f_{\nu_2}}{E_{\nu_1} - E_{\nu_2}}
F_{\nu\nu'}^{\nu_1\nu_2}
n_{\nu_1 \nu_2}^* {\bf (-q)}.
                             \ee
 where $F_{\nu\nu'}^{\nu_1\nu_2} $ are the matrix elements of 
the Fermi-liquid kernel:
        \be
F_{\nu \nu'}^{\nu_1 \nu_2} =
\varphi_{\alpha \alpha'}^{\alpha_1 \alpha_2}
\delta_{\sigma \sigma'} \delta_{\sigma_1 \sigma_2} +
\psi_{\alpha \alpha'}^{\alpha_1 \alpha_2} 
 ({\bf s}_{\sigma \sigma'} {\bf s}_{\sigma_1 \sigma_2} ).
         \ee
   Here, $ \alpha$ is the set of orbital quantum numbers, 
$ \sigma $ is the spin number, and $ \bf s$  is the operator 
of the electron spin.

The relations (1), (2) have to be complemented by the 
condition of electrical neutrality of the system :
   \be
\delta N ({\bf r}) +
e N {\mbox {div}} {\bf  u} ({\bf  r}) = 0 .
           \ee
  It follows from the equations (1),(2),(6) that:
   \be
{\mbox {curl}} \{ (1 - 4 \pi \chi){\bf  b} 
({\bf  r})\} = - 4 \pi {\mbox {curl}} 
\left \{\frac{\partial {\bf  M}}{\partial \zeta} 
\frac{N}{N_\zeta^*} {\mbox {div}}  
{\bf  u}({\bf  r}) \right \}.
    \ee

The set of simultaneous equations (1),(2),(6) was first
presented in previous works [1,6]. We use these equations 
to exclude {\bf b}({\bf r}) and to express the 
potential $ \varphi ({\bf  r}) $ in terms of the lattice 
displacement vector. As a result we  arrive at the expression 
for the electron force {\bf F}({\bf r}) acting upon the 
lattice under its displacement by the vector {\bf u}({\bf r}):
           \be
{\bf  F}({\bf  r}) =
\lambda_0 {\bf  b}_0 
\big( {\bf b}_0 \nabla (\nabla {\bf  u}
({\bf  r}))\big)  
  + \lambda \big [ {\bf  b}_0 \times
[ \nabla (\nabla {\bf  u}({\bf  r}))
\times {\bf  b}_0 ]\big ] .
              \ee
 Here, {\bf b}$_0 $ is the unit vector directed along { \bf 
B}.  This result (8) proves that the constants $ \lambda_0, \ 
\lambda_1$ represent electron contributions to the elastic 
constants corresponding to the deformation of the lattice 
along the external magnetic field $ (\lambda_0)$ and across 
this field $(\lambda).$ Within the adopted geometry these 
constants equal the electron terms in the compression elastic 
moduli $ c_{33}$ and $ c_{11} = c_{22}$ (in Voight notation). 
On the basis of the equations (1), (2),(6)  the 
expressions for these constants are derived [1,6]:
         \bea
\lambda_0 & = &\frac{N^2}{N_\zeta^*} ;
         \\   
\lambda & = &\lambda_0 \left ( 1 + \frac{4 \pi \chi_\zeta}
{1 - 4 \pi \chi_{||}} \right ) .
         \eea
 Here $\ds \chi_{||}= \frac{\partial M_z}{\partial B} + 
\frac{\partial M_z}{\partial \zeta} 
\frac{\partial \zeta}{\partial B}$ is the longitudinal 
part of the magnetic susceptibility;  
$ {\displaystyle \chi_\zeta = \frac {\partial
M_z}{\partial \zeta} \frac {\partial \zeta}{\partial B}}.$

As follows from Eq.(9), the quantity $ \lambda_0 $ coincides 
with the compression modulus of the electron liquid. The 
structure of the quantity $ \lambda $ is more complicated. 
Besides the electron compression contribution, $ \lambda $ 
also contains a contribution of a different origin.  This 
extra term appears due to the inhomogeneous magnetic field 
$  \bf b(r)$ caused with the lattice deformation. This field 
arises due to the change in the magnetization of electrons 
caused by the deformation. So, the appearance of the second 
term in the expression (10) is a manifestation of a 
magnetostriction effect.

\begin{figure}[t]
\begin{center}
\includegraphics[width=3.8cm,height=4.4cm]{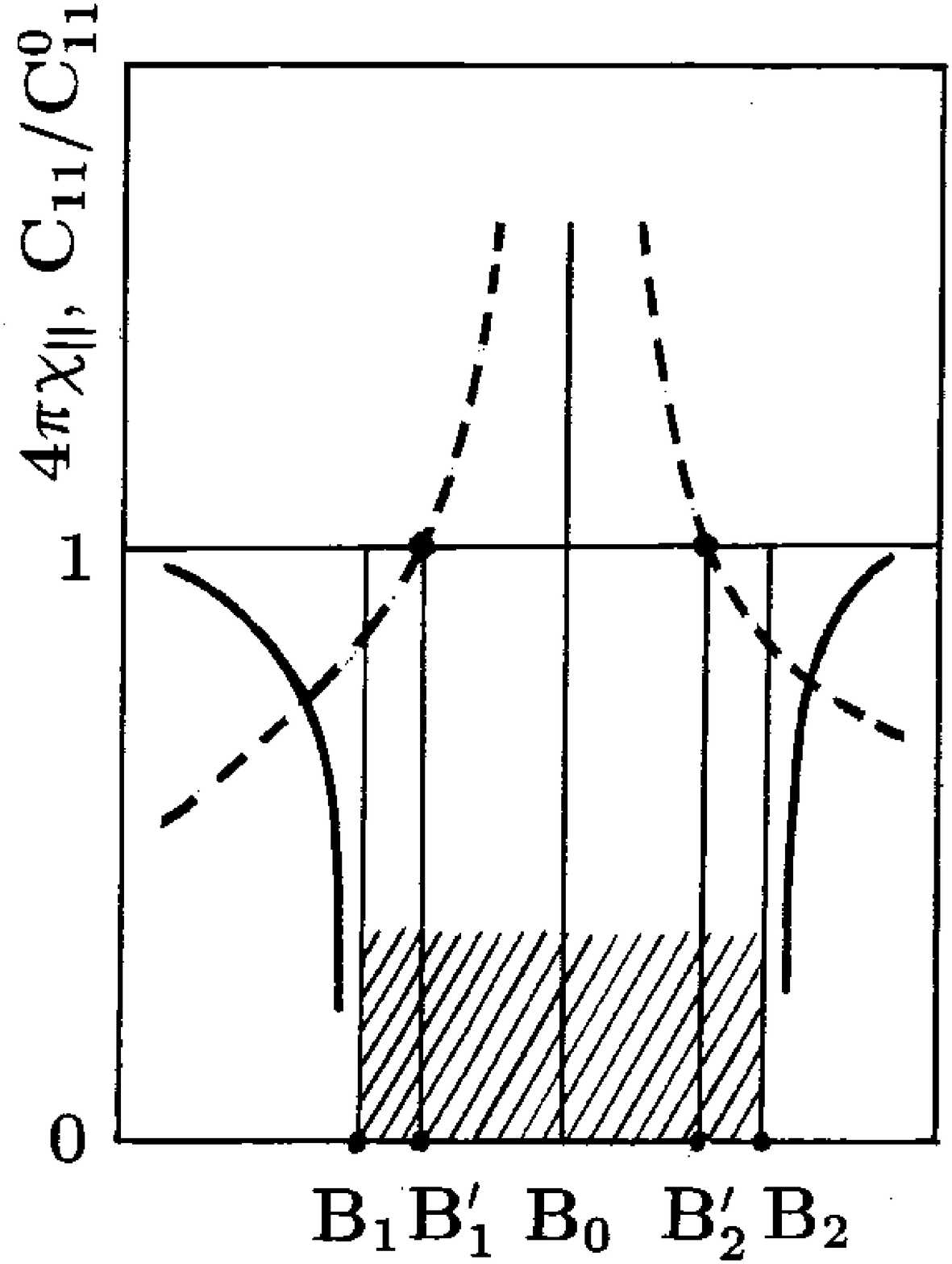}
\caption{
Schematic plot of the magnetic field dependence of
$4 \pi \chi_{||}$ (dashed line) and $ c_{11}/c_{11}^0$
(solid line) near a peak of quantum oscillations at $ B=B_0
 \ (T = 0).$ The range of magnetic fields corresponding
to structural $ (B_1 < B < B_2)$ and/or magnetic
$(B_1' < B < B_2')$ instability is section-lined.
}
\label{rateI}
\end{center}
\end{figure}

When the differential magnetic susceptibility  
enhances in the proximity of diamagnetic phase transition, 
the denominator of the second term in (10) can take 
on values close to zero.
Owing to this, the quantity $\lambda $ 
significantly grows in magnitude. For negative $ \lambda $
 this enhancement brings a noticeable decrease in the elastic
constant $ c_{11} $ (the latter is the sum of the "bare" 
elastic modulus $ c_{11}^0 $ and the electron contribution 
$ \lambda$). In other words, the expression (10) reveals a 
possibility for softening of a longitudinal acoustic
mode propagating perpendicularly to the magnetic field 
$ \bf B$. This possible softening 
arises due to magnetostriction, and it results from magnetic instability.
 
When a strong magnetic field is applied to a metal, this gives rise to
quantum oscillations in the electron DOS. The latter causes quantum
oscillations in observables including magnetic susceptibility $ \chi_{||}. $
At  low temperatures $ (\theta << 1) $ the magnitude of quantum oscillations
increases so much that the oscillating term could predominate at peaks of
oscillations. It was shown before (within the simple model of isotropic
electron liquid) that under such conditions both $ N_\zeta^* $ and $ 1 -
4 \pi \chi_{||} $ could go to zero near the oscillations peaks producing
magnetic instability of the metal and softening of some acoustic modes 
[1,5,6]. This is illustrated in the Fig. 1. Here, magnetic fields 
$ { B}_1' $ and $ { B}_2' $ label thresholds of the magnetic instability 
region, and the differential magnetic susceptibility diverges at these points. 
Singularity in the longitudinal susceptibility $ \chi_{||} $ discussed in the
earlier papers [1], appears significantly closer to the field $  B_0$
indicating the position of the oscillation peak. At the same time, the 
structural instability thresholds $ B_1$ and $ B_2 $ are located farther from
$ B_0 $ than $ B_1' $ and $ B_2' ,$ respectively, as shown in the Figure 1. 
However, these effects could be revealed in experiments
only at extremely low temperatures (of the order of 10 mK or lower). The
stringent requirements for temperatures explain why the softening of the
phonon modes at peaks of quantum oscillations was not observed so far.

It is known that under certain conditions local 
geometrical features of the FS could significantly affect 
the electronic response of the metal. This happens
when a dominating contribution to the response functions 
results from small "effective" segments of the FS, where 
"efficient" electrons are concentrated.
When the effective segments of the FS are nearly cylindrical 
or include locally flattened pieces, this gives a significant 
enhancement in the number of effective electrons, and can 
produce noticeable changes in the electronic response of 
the metal. The influence of locally flattened and neary 
cylindrical parts of the FS on the ultrasound attenuation 
rate, as well as on the surface impedance of a metall has
been analyzed before (see e.g. [8--10]). Quantum 
oscillations of the electron DOS in strong magnetic fields 
are specified  with contributions from effective cross-sections 
of the FS. Those are cross-sections with minimum 
and maximum sectional areas. Therefore the local geometry 
of the FS in the vicinities of these cross-sections has to 
affect both magnitude and shape of the oscillations [10]. 
This can influence anomalies of the elastic moduli
under the present study and  create much more favorable 
conditions for their observations  in  metals, as we show below.

We consider a metal whose FS is axially symmetric in the vicinity of an extremal
cross-section at $ p_z = p_0 $ with the area $ A_{ex}$. We assume the magnetic 
field $ \bf B $ to go along the symmetry axis. The general expression for the 
FS curvature near $ p_z = p_0 $ could  be written in the form:
   \be 
 K (x) = - \frac{1}{2} \ \frac{d^2 A/dx^2 }{p_m^2 A_{ex}}
   . \ee
 Here,  $ x = (p_z - p_0) / p_m; \ p_m $ is the maximum value of the quasimomentum component $ p_z $ at the FS. Now, we adopt the following approximation for the cross-sectional area around the extremal cross-section:
   \be 
 A(x) = A_{ex} ( 1 \pm b^2  x^{2l})
  \ee
 where $ b^2 $ is the dimensionless constant, $ (b^2 << 1) $ 
and the parameter $ l $ takes on values greater than $1. $

The very essence of the employed model (12) is that it describes an axially
symmetric FS whose curvature turns zero at $ p_z = p_0. $ The expression for
the curvature (11) showes that the approximation (12) is necessarily applicable
to each nearly cylindrical strip on any such FS. Otherwise the strip has a
nonzero curvature.  So we see that the model (12) gives the general expression
for cross-sectional area of any nearly cylindrical segment of an axially
symmetric FS.

Assuminig that the cyclotron quantum $ \hbar \Omega $ is small compared to 
$ \zeta \ ( \gamma^{-1} \equiv (\hbar \Omega /\zeta)^{1/2} << 1 ) $ and 
using the model (12) we arrive at the following expression for the contribution 
from the nearly-cylindrical cross-section to the electron DOS oscillations:
        \be
\Delta = \frac{\eta}{(\gamma)^{1/l}} \sum_{r=1}^\infty
\frac{(-1)^r}{r^{1/2l}} \psi_r (\theta) \cos \left 
(\pi r \gamma^2 \pm \frac{\pi}{4 l} \right ) \cos \left 
(\pi r \frac{\Omega_0}{\Omega} \right).     
    \ee
 Here, $\displaystyle{\psi_r (\theta) = \frac{r \theta}{\sinh r \theta}; 
\  \eta = \frac{\Gamma (1/2l)}{2l (b \sqrt\pi)^{1/l}};
\ \Gamma (x) }$ is the gamma function, and $ \hbar \Omega_0 $ is the spin 
splitting energy. Our formula (13) agrees with the results obtained for a 
precisely
cylindrical FS/and a Fermi circle in a two-dimensional conductor (see e.g.
[11,12]). We arrive at the corresponding results within the limit $ l \to
\infty. $
In general case we can treat $ "l "$ as a phenomenological parameter included
in the model (12).  Actual values of $ "l"$ could be discovered in
experiments where the FS local geometry is revealed. This is the only
trustworthy way to estimate the above parameter for a particular metal.
First  principle calculations are not accurate enough to produce reliable
results on fine geometrical features of FSs although their shapes as a whole
are well known.

 Within the isotropic model the cross sectional area is described with the 
expression (12) where $ l = b^2 = 1 , \ p_0 = 0, $ and $ p_m $ is the 
radius of the Fermi sphere in quasimomemta space. In this particular case 
the oscillating function $ \Delta $ takes on a well-known form:
        \be
\Delta = \frac{1}{\gamma}
\sum\limits_{r=1}^\infty
\frac{(-1)^r}{\sqrt r} \psi_r(\theta)
\cos\left(\pi r \gamma^2 - \frac{\pi}{4}\right)
\cos\left(\pi r\frac{\Omega_0}{\Omega}\right).
         \ee

Oscillations described by (13) and (14) differ in phase 
as well as amplitude. The amplitude of usual oscillations 
given by (14) is of the order of $ \gamma^{-1} \theta^{-1/2},
 $ while (13) gives a magnitude of the order of 
$ \gamma^{-1/l} \theta^{(1-2l)/2l}.$ Therefore, the amplitude
of oscillations related to the extremal section of zero 
curvature is approximately $ (\gamma^{-1} 
\theta^{1/2})^{(1-l)/l} $ times greater than that of the 
usual quantum  oscillations. As a result,  the contribution due 
to an extremal section with zero curvature can be considerably 
(more that tenfold) greater 
than contributions due to other extremal sections, and 
the function $\Delta $ can reach values of the order of 
unity at peaks of oscillations even at $ \theta \sim 1 . $ 
On these grounds we conclude that the most favorable conditions 
for observation of softening of elastic moduli at peaks of 
quantum oscillations occur in
metals whose FSs include nearly cylindrical segments.
We consider such FSs in further analysis.
                                                                                
Assuming $ \gamma << 1 $ we replace matrix elements included in the 
Fermi-liquid kernel with their semiclassical analogs $ \varphi \bf (p, p') $  
and $ \psi \bf (p, p') $ which depend on quasimomenta of interacting 
conduction electrons $ \bf p $ and $ \bf p'. $
For axially symmetric FSs the Fermi-liquid functions take the form:
       \be 
\varphi ({\bf p, p}') = \varphi_{00} + p_z p_z' \varphi_{01}
+ ({\bf p_\perp p'_\perp})  (\varphi_{10} + p_z p_z'
 \varphi_{11});
      \ee
   \be       
\psi ({\bf p, p}') = \psi_{00} + p_z p_z' \psi_{01}
+ ({\bf p_\perp p'_\perp})  (\psi_{10} + p_z p_z' \psi_{11}).
      \ee
                                                                                
To make our analysis more thorough we include  deformation terms in  
the force (8) exerted by the conduction electrons on the lattice.
We present the components of the deformation potential, as:
        \be
\Lambda_{\alpha \beta} ({\bf p}) = \Lambda_{\alpha \beta} + 
G \Pi_{\alpha \beta} (\bf p),
         \ee
 where $ \Lambda_{\alpha \beta} $ is a tensor whose elements 
do not depend on $ {\bf p}, \ G $ is a dimensionless 
constant, and $ \Pi_{\alpha \beta} $ is the tensor of the 
electron momentum flux density.

The adopted approximations lead to the following expressions 
for the corrections to the elastic constants:
        \be
\tilde c_{11} = \tilde c_{22} = - \frac{N^2}{g}  \ 
\frac{[1 + G/(1 + A_2)]^2
\Delta}{1 + (1+ W - 4 \pi \chi_0 \gamma^4) \Delta}; 
       \ee
        \be
\tilde c_{12} = - \frac{N^2}{g}  \left (\frac{G}{1 + A_2} 
\right )^2 \frac{
\Delta}{1 + (1+ W - 4 \pi \chi_0 \gamma^4) \Delta}.
       \ee
   where $ g $ is the electron DOS in the absence of the
external magnetic field,
      \be 
 \tilde c_{33} = - \frac{N^2}{g} 
\frac{[1 + G (1 + A_2)^2] \Delta}{1 + (1 + W) \Delta}
   \ee
 $ \chi_0 $ is related to the Landau
diamagnetic susceptibility (the latter equals $ - \frac{1}{3}
\chi_0 $); and the constant $W$ originates
from the Fermi-liquid interaction:
 \be
W = \frac{B_0}{1 + B_0} +\frac{B_2}{1 + B_2} + 
\frac{A_2}{1 + A_2}.
    \ee
  Here,  dimensionless coefficients $B_0,  B_2  $ and 
$  A_2  $ are related to the Fermi-liquid parameters 
$ \psi_{00}, \psi_{11}, $ and  $\varphi_{11} $ 
as follows:
      \bea
B_0 & = &\frac{1}{(2 \pi h)^3} \int \psi_{00} S (p_z) d p_z;
 \nn \\
B_2 & = & \frac{1}{(2 \pi h)^3} \int \psi_{11} S (p_z) p_z d p_z;
  \nn\\
A_2 &= & \frac{1}{(2 \pi h)^3} \int \varphi_{11} S (p_z)p_z d p_z.
     \eea                                                                       
 It is noteworthy to mention that the shear modulus $ c_{12} $ is also affected 
due to magnetostriction. This was not discovered in the early analysis of [1] 
where the deformation contributions into the electron force were omitted.
As for elastic constant  $  c_{33} $, it also can be reduced
at the peaks of low temperature quantum oscillations but this
effect is not related to magnetic instability. The
expression for  $\tilde c_{33} $ (20) does not include the
contribution arising due to magnetostriction. The possible
softening of $ c_{33} $ is directly connected with the
behavior of the electron DOS under strong magnetic fields at
low temperatures. Such instability was predicted before  [7].
We have to remark here that the interaction among electrons
significantly influences all above effects. The value of the
constant $ W $  which accumulates effects of electron-electron
interactions within the framework of the Fermi liquid theory,
could significantly influence the temperature range where
both magnetic and lattice instabilities  occur.

The  function $ \Delta $ describing quantum oscillations is 
given by Eq. (13) and the amplitude of oscillations may 
become comparable with unity at moderately low temperatures
provided that the FS shape reveals a fair proximity to a 
cylinder near the extremal cross section. 
For example, at $ l = 3, \ h \Omega  /\zeta \sim 10^{-3} ,$ 
and $ B \sim 10 T $, the condition $\gamma ^{-1 / l} 
\theta^{(1 - 2l)/2 l} \sim 1 $ could  be satisfied at 
temperatures of the order of 1 K. 
In usual metals $ N \sim 10^{21} \div 10^{22} $cm$^{-3},$ and the term 
$ 4 \pi \chi_0 \gamma^4$ in the denominator of (18) can 
take on values of the order of 10  at $ \gamma^2 \sim 10^3.$ 
Also, we introduce the quantity
$ L^2 = N^2 (1 + G /(1+A_2))^2 \big/g c^0, $ where 
$c^0$ is the relevant "bare" elastic constant. The ratio 
$ N^2/gc^o $ in typical metals is rather small $(N^2 /gc^0 
\sim 10^{-2} \div 10^{-1}).$ However, the values taken on 
by $L^2$ could be noticeably greater than that due to the 
deformation constant $G.$ The latter mostly accepts values 
of the order of unity. So, we have grounds to expect the 
values of $ L^2 $ to be of the order of $ 10^{-1} $ rather 
than $ 10^{-2}. $

To proceed in the analysis of the experimental feasibility
of the effect, we numerically evaluate the decrease in the elastic 
constant $ c_{11} $ using our result (18) and the above estimations of the 
parameters included there. The results are shown in the Fig. 2. We see 
that close enough proximity in the shape of an effective strip on the FS 
to a cylinder gives rise to the structural instability near peaks of 
oscillations at $ \theta = 1. $ Also, it is demonstrated that the effect 
is washed out due to the further rise in temperature.

\begin{figure}[t]

\begin{center}
\includegraphics[width=7.2cm,height=4.0cm]{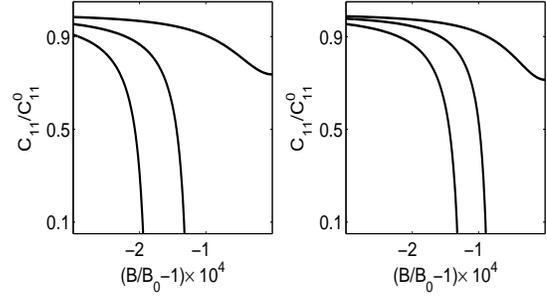}
\caption{
Magnetic feld dependencies of the elastic constant $C_{11} $ near
the diamagnetic phase transitions. The curves plotted for:  $\  \theta =1,$
the parameter $ l $ takes on values $8,4,2 $ from the left to the right  (left panel); 
$\ l = 4,  \ \theta $ takes on values $ 1,2, 3 $ from the left to the right (right panel).
For all curves $\ N = 10^{21} $cm$^{-3}, \ \gamma^2 = 10^3, \ B_0 = 10 T. $
}
\label{rateI}
\end{center}
\end{figure}

Electronic contributions to the velocity of ultrasound waves 
propagating in metals are simply related to the elastic 
constants. It follows from our results that the velocity of a 
longitudinal sound could  depend on the 
direction of its propagation. In the vicinity
of the Condon instability the velocity of sound propagating
perpendicularly to the magnetic field $ \bf B $ could be
 noticeably reduced compared to the velocity of sound 
propagating along $ \bf B. $ Again, we may expect this 
effect to appear in metals whose FSs include nearly 
cylindrical segments.

Finally, it is known that  the effect of
magnetostriction can cause softening of some phonon modes in metals near 
the Condon magnetic instability. This effect can appear even in 
an isotropic metal  [1,6]. The point of the present work is that the 
effects could be significantly
strengthened when the immediate vicinities of some extremal cross-sections
of the FS are nearly cylindrical in shape, so that the FS curvature turns
zero at these cross-sections [13].there is an 
Real metals mostly have nonspherical and complicated 
in shape FSs.  At present  the main geometric characteristics
 of the FSs, such as their connectivity, are there is an well studied. On 
the contrary,  local geometric features of the FSs has not 
been investigated in detail so far. However, there is an 
experimental evidence that "necks" connecting quasispherical 
pieces of the FS of copper include nearly cylindrical belts 
[11]. When the magnetic field is directed along the axis of 
a "neck" (for instance, along the [111] direction in the 
quasimomenta space), the extremal cross section of the 
"neck" could be expected to run along the nearly cylindrical 
strip where the FS curvature turns zero. It is also likely 
that the FS of gold possesses the same geometrical features 
for it closely resembles that of copper. Another kind of 
materials where we can expect the low temperature softening 
of phonon modes to be manifested includes some layered 
structures with metallic-type conductivity 
(e.g. $\alpha-(BEDT-TTF)_2MHg(SCN)_4 $ group of organic 
metals). Fermi surfaces of these materials are sets of 
rippled cylinders, isolated or connected by links [14]. 
Based on the experiments on cyclotron resonance in these 
organic metals [15], it was shown that the cylinders could 
have nearly cylindrical strips [16]. For all above listed 
substances we can expect the effect to be revealed at 
reasonably low temperatures $ (\sim 1K)$ and reasonably 
strong magnetic fields $ (1 \div 10 T).$

The effect is expected to be very sensitive to the geometry 
of an experiment for extremal cross sections of the FS run 
along nearly cylindrical belts (if any) only at certain 
directions of the magnetic field. When the magnetic field is 
tilted away from such direction, the extremal cross section 
slips from the nearly cylindrical piece of the FS. This does 
not cancel possibilities of the effect in principle but 
makes requirements for temperatures dramatically stringent, 
as it was discussed above. The present analysis is carried 
out within the model of axially symmetric FS. The obtained 
results could be applied to actual metals when the magnetic 
field is directed in parallel with a high order axis of 
symmetry of the crystalline lattice of a metal. Otherwise, 
it is very difficult to separate out electron contributions 
to particular elastic constants.
In summary, the effect of softening of phonon modes at peaks 
of quantum oscillations in metals could be expected to be 
observed in experiments. The most favorable conditions for 
the effect exist in metals whose FSs include nearly 
cylindrical pieces. The efffect could be revealed in these 
metals for some particular directions of the external 
magnetic field providing that an extremal cross section 
belongs  to a quasicylindrical strip on the FS.

{\it  Acknowledgments:}
I thank G.M.Zimbovsky for help with the manuscript.
This work was supported  in part by NSF Advance  program SBE-0123654.

\end{document}